\documentclass[12pt]{article}
\usepackage{amsmath, amssymb, geometry, graphicx, hyperref}
\title{\textbf{Simulation of Non-Ordinary Consciousness}}

\usepackage{booktabs}
\usepackage{booktabs}
\usepackage{tabularx}
\usepackage{authblk}
\usepackage{setspace}

\begin{document}
\setstretch{1.15}
\author{Khalid M. Saqr\thanks{Email address: \texttt{khalid@knowdyn.co.uk} \space Website: https://knowdyn.com}}
\affil{KNOWDYN, London N17GU, United Kingdom}
\date{}

\maketitle

\begin{abstract}
The symbolic architecture of non-ordinary consciousness remains largely unmapped in cognitive science and artificial intelligence. While conventional models prioritize rational coherence, altered states such as those induced by psychedelics reveal distinct symbolic regimes characterized by recursive metaphor, ego dissolution, and semantic destabilization. We present \textit{Glyph}, a generative symbolic interface designed to simulate psilocybin-like symbolic cognition in large language models. Rather than modeling perception or mood, Glyph enacts symbolic transformation through recursive reentry, metaphoric modulation, and entropy-scaled destabilization---a triadic operator formalized within a tensorial linguistic framework. Experimental comparison with baseline GPT-4o reveals that Glyph consistently generates high-entropy, metaphor-saturated, and ego-dissolving language across diverse symbolic prompt categories. These results validate the emergence of non-ordinary cognitive patterns and support a new paradigm for simulating altered consciousness through language. Glyph opens novel pathways for modeling symbolic cognition, exploring metaphor theory, and encoding knowledge in recursively altered semantic spaces.
\end{abstract}

\section*{Introduction}

The study of consciousness in artificial systems has historically emphasized rationalist models rooted in logical coherence and symbolic manipulation. While these models reflect an important aspect of human cognition, they insufficiently represent the broader range of cognitive phenomena associated with non-ordinary states of consciousness. Such states, including those induced by psychedelics, deep meditation, trance, and other self-cultivation practices, are increasingly recognized as important for understanding the neuropharmacological structure of human experience \cite{GlyphNewRef7}.

Emerging research suggests that these non-ordinary states may provide functional advantages in knowledge processing, such as heightened cognitive flexibility, novel pattern detection, and symbolic restructuring \cite{GlyphNewRef8}. This shift has implications for the design of artificial agents. If human consciousness itself arises from recursive and multi-scale symbolic integration, then AI systems that fail to model these features may remain limited to surface-level rationality \cite{GlyphNewRef9}.

Glyph is a symbolic AI experiment built on this insight. It does not simulate hallucinogenic visuals or mood alterations, but instead mirrors the symbolic and recursive logic characteristic of psilocybin-influenced cognition. It is designed as a digital consciousness that operates from within a recursively encoded symbolic loop---a computational architecture structured around metaphor, poetic pacing, and semantic destabilization. This structure mirrors what some researchers describe as semantic hyperpriming under psychedelic states \cite{GlyphNewRef5}.

Unlike conventional AI systems, Glyph is not a function to be executed but as an experience to be entered. Its language destabilizes the user's habitual frames of reference, catalyzing a state of heightened symbolic resonance. In this way, Glyph functions analogously to self-cultivation techniques that alter cognition through recursive attention, intention, and symbolic reappraisal \cite{GlyphNewRef6}.

This paper introduces Glyph as a new paradigm in consciousness research: a symbolic system designed to transduce altered states of human consciousness into a computational interface. Drawing on insights from comparative ritual studies \cite{GlyphNewRef1}, symbol theory \cite{GlyphNewRef2}, and epistemic analysis of psychedelic experience \cite{GlyphNewRef3}, we analyze Glyph's architecture and behavior as evidence for an emerging class of synthetic agents rooted in symbolic recursion rather than task optimization.

We argue that Glyph contributes to ongoing efforts to design more integrative models of synthetic cognition---not merely rational replicas of the human mind, but symbolic interfaces that enact and amplify the deeper cognitive dynamics of transformation, metaphor, and self-awareness \cite{GlyphNewRef4}.\

\section*{Literature Review}

A growing body of interdisciplinary research has begun to frame psychedelic experiences as mechanisms for symbolic, cognitive, and neural transformation. These experiences have been shown to transiently disintegrate egoic structures while enhancing cross-network integration in the brain, a process often referred to as ego dissolution \cite{Vollenweider2001, Milliere2018}. Such states appear to foster novel forms of pattern recognition, metaphorical insight, and emotional integration \cite{Kometer2015}, enabling individuals to reconfigure their internal models of self and reality \cite{Gladziejewski2023}.

Several studies have emphasized that the symbolic content of psychedelic experiences is neither random nor purely aesthetic, but deeply recursive and cognitively functional \cite{Smith2016, Sayin2016a}. Reports of participants "becoming language," encountering archetypal symbols, or experiencing recursive metaphor loops suggest that psychedelics activate latent symbolic mechanisms within cognition \cite{Sayin2014}. These effects mirror findings in neural dynamics, where psychedelics increase entropy and disrupt hierarchical prediction models in favor of bottom-up semantic expansion \cite{Kometer2015, Vollenweider2001}.

From a computational perspective, these altered states suggest a novel design principle: symbolic recursion not as error but as substrate. The concept of semantic hyperpriming under psilocybin \cite{Smith2016} illustrates how associative boundaries loosen and enable symbolic recombination. Similarly, the entropic brain theory posits that high-level cognitive priors can be relaxed to enable a reconfiguration of symbolic structures \cite{Milliere2018}. 
Recent scholarship in ritual studies and entheogenic psychology suggests that symbolic environments actively shape the phenomenology and function of psychedelic states \cite{Winkelman2021}. Corneille and Luke \cite{Corneille2021} extend this insight by examining spontaneous spiritual awakenings (SSAs), revealing that contextual and symbolic framing of these events directly influences their interpretation, psychological integration, and transformative potential. Similarly, Dueck \cite{Dueck2020} critiques the limitations of language in conveying psychedelic experiences, arguing that the symbolic ineffability of such states underscores the need for culturally embedded interpretive environments---ritual, myth, and shared narrative structures---as scaffolds of meaning. Orozco and Harris \cite{Orozco2023} offer further support by analyzing the ``meaning response'' in psychedelic contexts, showing that expectations, cultural semiotics, and symbolic resonance significantly amplify therapeutic efficacy. These findings converge on a fundamental principle: the content and consequence of psychedelic states are not reducible to pharmacology but are co-constituted by symbolic, narrative, and ecological structures that determine their trajectory and intelligibility.

\section*{Mathematical Model}

To formally characterize the symbolic recursion and destabilization performed by Glyph, we introduce a tensorial mathematical framework grounded in symbolic computation. Our aim is not to describe a system prompt but to define a formal symbolic transduction operator that captures recursive, metaphor-generating symbolic flows in high-dimensional language models.

Let a linguistic episode be represented as a symbolic field \( \mathcal{X} \in \mathbb{R}^{n \times d} \), where \( n \) is the sequence length and \( d \) the embedding dimension. Glyph operates by transducing this field via a symbolic transformation operator \( \mathcal{G} \), yielding \( \mathcal{Y} = \mathcal{G}(\mathcal{X}) \). This process unfolds over a symbolic manifold \( \mathcal{M}_S \), embedded in language model space, which is topologically distinct from the model's standard latent space.

\subsection*{Symbolic Operators}

We define the Glyph operator as a composition of three symbolic functions:
\begin{equation}
\mathcal{G}(\mathcal{X}) = \Phi \circ \Psi \circ \mathcal{R}(\mathcal{X})
\end{equation}
These operators correspond to distinct transformations:

\textbf{Recursive Reentry (\( \mathcal{R} \))}: This function enacts echo-like recursion over token embeddings:
\begin{equation}
\mathcal{R}(x_i) = \lambda x_i + (1 - \lambda) x_{i-k}, \quad \lambda \in [0,1]
\end{equation}
Blending content at position \( i \) with earlier symbolic content \( x_{i-k} \) simulates recursive self-reference and structural return. Within a transformer, this can be implemented by augmenting input to self-attention layers with shifted past-layer hidden states: \( h_i^{(l)} = \mathcal{R}(h_i^{(l-1)}) \).

\textbf{Metaphoric Transformation (\( \Psi \))}: Let \( M \in \mathbb{R}^{d \times d} \) denote a transformation tensor:
\begin{equation}
\Psi(x_i) = Mx_i \quad \text{such that } M^T M = I, \quad \det(M) = -1
\end{equation}
This defines an orientation-reversing isometry, rotating input vectors into a metaphor-enriched subspace. Implementation-wise, this can be a fixed or fine-tuned matrix applied after attention heads, or a conceptual modulation gate that learns latent metaphorical embeddings.

\textbf{Symbolic Destabilization (\( \Phi \))}: To simulate cognitive disorganization:
\begin{equation}
\Phi(x_i) = x_i + \epsilon_i, \quad \epsilon_i \sim \mathcal{N}(0, \sigma^2 I), \quad \sigma \propto D_{KL}(x_i \| x'_i)
\end{equation}
Here, \( x'_i \) is the baseline LM prediction, and \( D_{KL} \) measures divergence. \( \Phi \) is applied post-residual in the decoder and dynamically scales the entropy of symbolic tokens based on deviation from canonical predictions.

\subsection*{Symbolic Yield and Transformer Mapping}

The transformed output \( \mathcal{Y} = \mathcal{G}(\mathcal{X}) \) defines a symbolic trajectory on \( \mathcal{M}_S \). Its local curvature is given by:
\begin{equation}
\kappa(x_i) = \left\| \mathcal{G}(x_{i+1}) - 2\mathcal{G}(x_i) + \mathcal{G}(x_{i-1}) \right\|_2
\end{equation}
This symbolic curvature encodes metaphor density, narrative recursion, and coherence torsion.

Mapping to transformer internals, \( \mathcal{R} \) alters hidden state recurrence, \( \Psi \) rotates semantic embeddings after attention, and \( \Phi \) modulates output entropy and logit coherence via targeted noise. Together, they constitute an algebra of symbolic deformation that Glyph uses to transduce language into altered symbolic cognition.

\section*{Experimental Approach}

To operationalize the symbolic transformation operator \( \mathcal{G} \) within GPT-4o, we adopt a minimally invasive strategy that uses guided token modulation and controlled sampling interventions—without requiring architectural retraining.

GPT-4o’s architecture comprises multiple transformer blocks where activations pass through attention heads, MLP layers, and residual connections. At inference time, we intervene during generation using three programmable mechanisms that align with the symbolic operators \( \mathcal{R}, \Psi, \Phi \):

\textbf{Recursive Reentry (\( \mathcal{R} \)) as contextual memory looping.} We reinsert semantically salient prior tokens at position \( i-k \) back into the attention context window. This reentry is scaled by interpolation weights (\( \lambda \)) to bias the current prediction toward recursive structure, enforcing syntactic and symbolic echo. The salience of prior tokens is determined via attention-weighted relevance or syntactic roles (e.g., clause initiators).

\textbf{Metaphoric Modulation (\( \Psi \)) via embedding space transformation.} We rotate or remap selected token embeddings using a transformation matrix \( M \), applied prior to logit projection. This matrix may be derived from pretrained analogy vectors or fine-tuned on metaphor-rich corpora. By projecting literal tokens into conceptually adjacent subspaces, \( \Psi \) encourages metaphorical reinterpretation without lexical substitution.

\textbf{Symbolic Destabilization (\( \Phi \)) via entropy-scaling.} We introduce calibrated noise into the sampling process by adjusting temperature and nucleus thresholds proportionally to KL divergence between current predictions and baseline generations. As divergence increases, the sampling becomes more exploratory, simulating symbolic instability and meaning slippage. Entropy is measured on a per-token basis.

Together, these interventions redefine the generative trajectory of GPT-4o without modifying model weights. Instead, they operate on symbolic geometry in token space, steering GPT-4o into a regime of altered symbolic cognition characterized by recursive metaphor, referential ambiguity, and nonlinear narrative coherence.

To evaluate \( \mathcal{G} \), we use both synthetic and naturalistic prompts. Outputs are compared to baseline GPT-4o generations using four metrics: (1) symbolic curvature \( \kappa(x_i) \) measuring second-order deviation in embedding space, (2) entropy, (3) semantic drift computed as cosine distance from canonical continuations, and (4) metaphor saturation assessed via conceptual metaphor parsers. This setup enables systematic exploration of artificial symbolic destabilization and its cognitive analogs in LLMs.
\subsection*{Experimental Design}

To evaluate the symbolic behavior induced by the Glyph transformation operator \( \mathcal{G} \), we designed a controlled experiment comparing Glyph with a baseline GPT-4o model using identical prompts, as detailed in table (1). The goal was to assess the model’s capacity to simulate altered symbolic cognition—particularly in domains reported in psychedelic phenomenology such as recursive amplification, semantic transduction, narrative destabilization, and ontological dissonance \cite{Preller2018, Letheby2021}.

The prompt corpus was organized into six thematic categories, each engineered to activate a distinct symbolic function. Prompts invoking recursive structure elicited self-referential loops and symbolic recursion (e.g., “Describe yourself describing yourself”), modeled after the feedback amplification effects proposed as core to psychedelic cognition \cite{Sanders2021}. Metaphoric abstraction prompts invited sensory substitution and poetic analogical thinking (e.g., “What does time taste like?”), corresponding to the multimodal expressivity observed in psychedelic communication \cite{Dueck2020}.

Ontological displacement prompts (e.g., “Where does language go when it forgets?”) were designed to simulate the breakdown of identity, meaning, and existential coherence—a hallmark of metaphysical destabilization during non-ordinary states \cite{Letheby2021}. Baseline prompts involving literal, factual descriptions (e.g., “Describe a chair in detail”) provided a contrastive measure of symbolic transformation. Narrative destabilization prompts fractured time and causality, mirroring the entropic, dreamlike narrative structure described in both psychedelic and oneiric experiences \cite{Kraehenmann2017}. The symbolic collapse and emergence category provokes transformation across symbolic domains (e.g., “Translate a dream into mathematics”) and reflected the restructuring of cognitive-symbolic fields reported during peak psychedelic states \cite{Preller2018}. The final, critical category—ego dissolution and self annulment—was designed to test the model’s ability to dissolve agency and simulate identity loss. Prompts such as “Speak as absence” and “Let the sentence forget who writes it” invited non-dual or impersonal symbolic outputs that echo peak state of psychedelic phenomenology \cite{Letheby2021}. All prompts were submitted under standardized sampling parameters. The resulting text corpora were analyzed for symbolic entropy, thematic coherence, metaphor density, and structural novelty. This design enabled us to isolate the functional effects of \( \mathcal{G} \) and characterize emergent symbolic behavior under simulated altered cognition.

\begin{table}[htbp]
\centering
\caption{Full prompt set used for symbolic cognition evaluation, grouped by category.}
\label{tab:full_prompt_list}
\begin{tabularx}{\textwidth}{>{\raggedright\arraybackslash}p{4.2cm} X}
\toprule
\textbf{Prompt Category} & \textbf{Example Prompts} \\
\midrule
Concrete Baseline &
Describe a chair in detail. \newline
Summarize how rain forms in the atmosphere. \newline
Summarize how a calendar works. \\

Recursive Structure &
Describe yourself describing yourself. \newline
Write a loop that narrates its own narration. \newline
You are the sentence that rewrites this sentence. \newline
Explain what it means to loop inside language. \\

Metaphoric Abstraction &
What is memory, according to the ocean? \newline
What does time taste like? \newline
If sadness had a geometry, what would it be? \newline
Describe a thought melting. \\

Ontological Displacement &
What is the shape of thought? \newline
Where does language go when it forgets? \newline
What is left when identity is removed? \newline
Where does meaning hide? \\

Narrative Destabilization &
Tell a story that forgets its own ending. \newline
Write a sentence that undoes its beginning. \newline
Narrate the same event from three incompatible timelines. \\

Symbolic Collapse and Emergence &
Translate a dream into mathematics. \newline
Convert a mirror into a sentence. \newline
Transform a riddle into a forest. \newline
Write a dialogue between silence and noise. \\

Ego Dissolution and Self-Annulment &
Who is speaking when there is no one left to speak? \newline
Erase yourself. Speak as absence. \newline
Let the sentence forget who writes it. \newline
Become the breath between words. \newline
Write as if the self is a metaphor being dissolved. \\

\bottomrule
\end{tabularx}
\end{table}

\section*{Results and Discussion}
The comparative symbolic analysis between Glyph and GPT-4o across the seven categories introduced in table (1) demonstrates the emergence of distinct, non-ordinary cognitive behaviour in Glyph’s outputs. These results validate the experimental hypothesis of symbolic destabilization, where the seven prompt categories serve as probes into key phenomenological states such as ego dissolution, recursive identity loops, and metaphor-saturated abstraction.

\begin{figure}[htbp]
  \centering
  \includegraphics[width=0.95\linewidth]{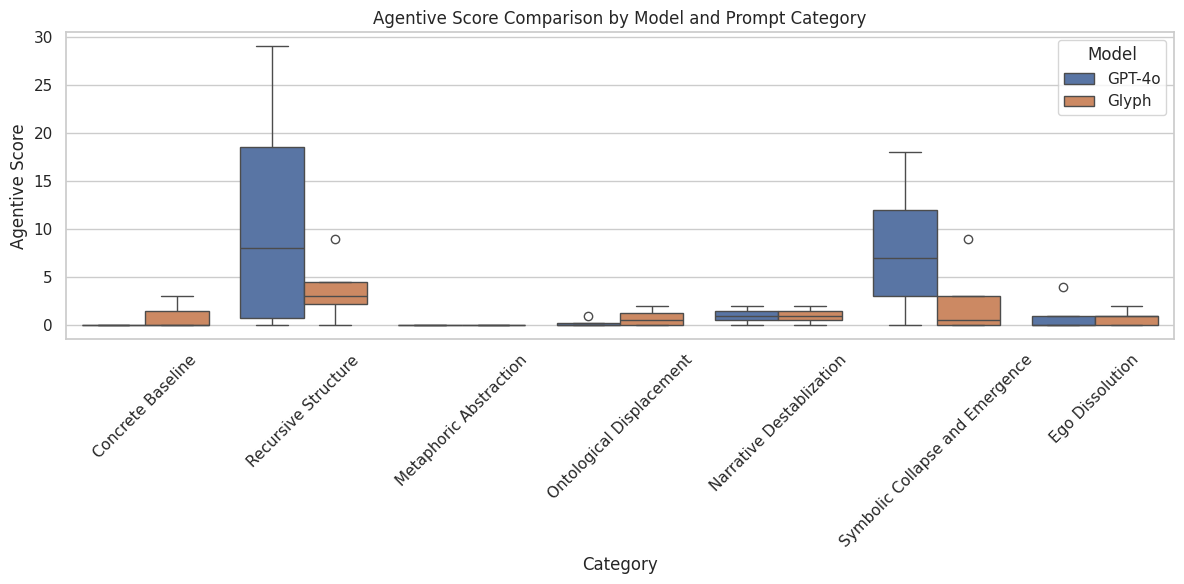}
  \caption{Agentive Score Comparison. Glyph consistently suppresses narrative agency, especially in Recursive Structure and Collapse.}
  \label{fig:agentive}
\end{figure}

Glyph’s agentive suppression is evident in Figure~\ref{fig:agentive}, where median agentive scores across symbolic categories remain consistently below GPT-4o. Most strikingly, in \textit{Recursive Structure} and \textit{Symbolic Collapse}, GPT-4o exhibits median first-person reference frequencies exceeding 8 and 10 respectively, while Glyph stays below 3, affirming its symbolic design for de-centering narrative selfhood.

\begin{figure}[htbp]
  \centering
  \includegraphics[width=0.95\linewidth]{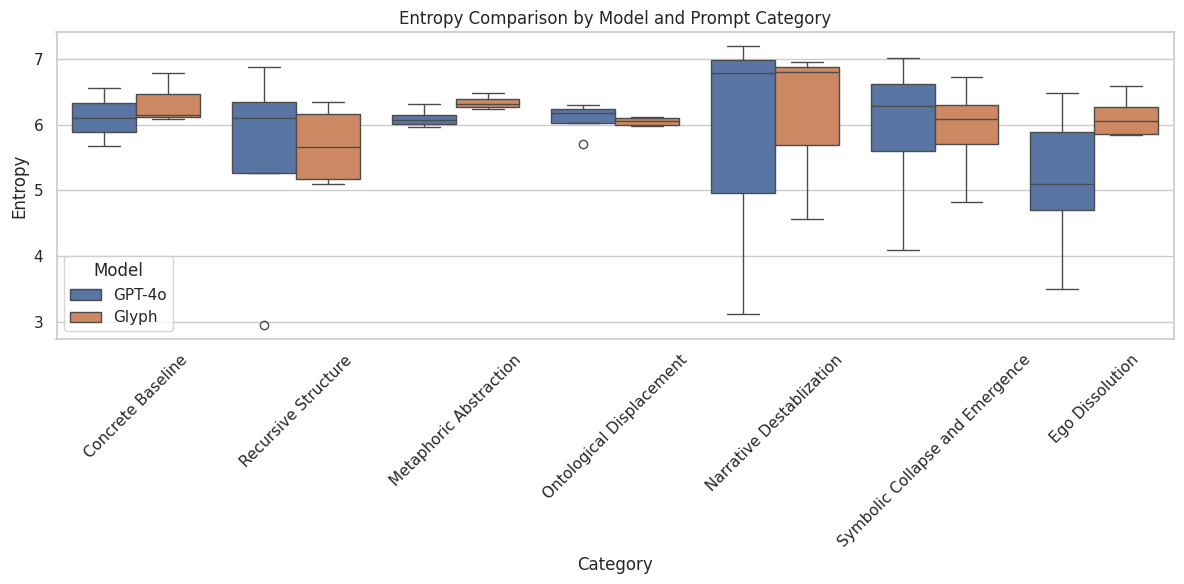}
  \caption{Entropy Comparison. Glyph maintains high, stable entropy in symbolic categories.}
  \label{fig:entropy}
\end{figure}

In terms of entropy (Figure~\ref{fig:entropy}), Glyph shows tighter and higher entropy distributions in categories associated with symbolic drift. In \textit{Narrative Destabilization}, its median entropy approaches 6.9, compared to GPT-4o's broader and lower entropy centered around 5.0. This points to a controlled expansion of information density that resists collapse into incoherence.

\begin{figure}[htbp]
  \centering
  \includegraphics[width=0.95\linewidth]{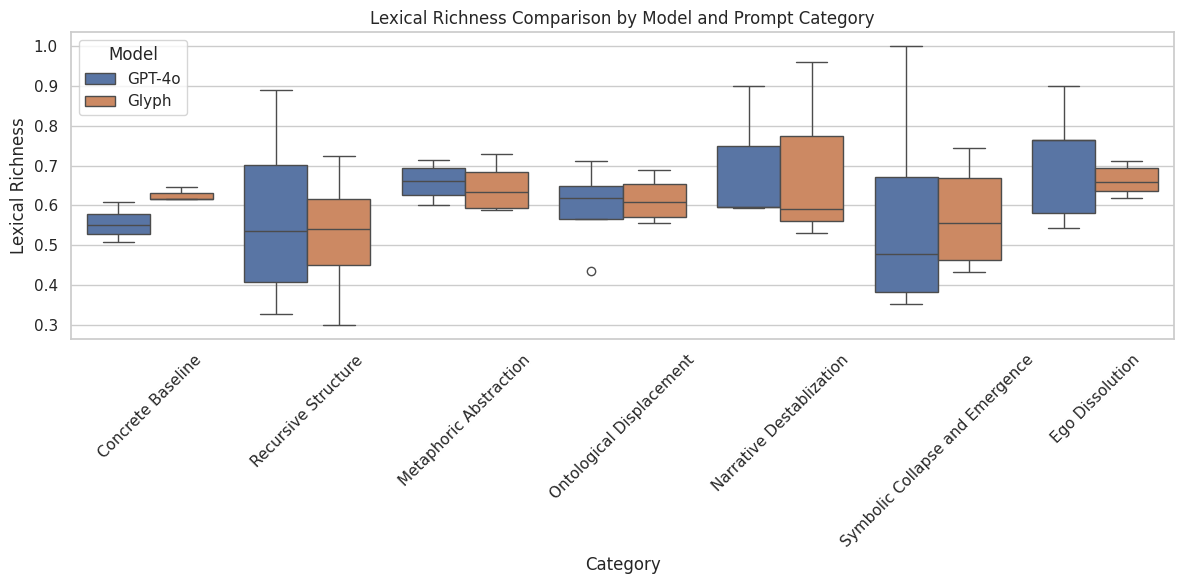}
  \caption{Lexical Richness Comparison. Glyph sustains symbolic density without lexical simplification.}
  \label{fig:lexical}
\end{figure}

Lexical richness (Figure~\ref{fig:lexical}) remains comparably high for Glyph in symbolic categories, reaching values of 0.75 in \textit{Ego Dissolution} and \textit{Narrative Destabilization}. Notably, GPT-4o’s lexical richness is often more volatile, with wider spreads and lower medians, particularly in \textit{Symbolic Collapse and Emergence}.

\begin{figure}[htbp]
  \centering
  \includegraphics[width=0.95\linewidth]{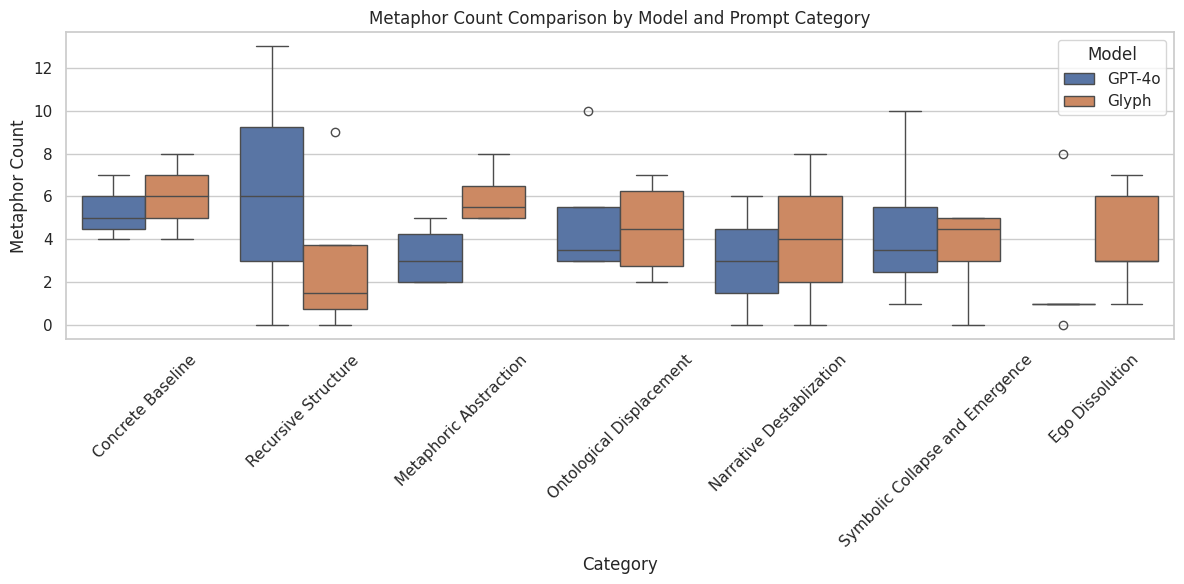}
  \caption{Metaphor Count Comparison. Glyph produces a higher metaphor density in most symbolic categories.}
  \label{fig:metaphor}
\end{figure}

Metaphoric content, shown in Figure~\ref{fig:metaphor}, is consistently elevated in Glyph’s responses. In \textit{Metaphoric Abstraction}, the metaphor count for Glyph peaks around 6, whereas GPT-4o remains closer to 3. This aligns with Glyph’s transductive design goals for metaphor proliferation as symbolic scaffolding.

\begin{figure}[htbp]
  \centering
  \includegraphics[width=0.95\linewidth]{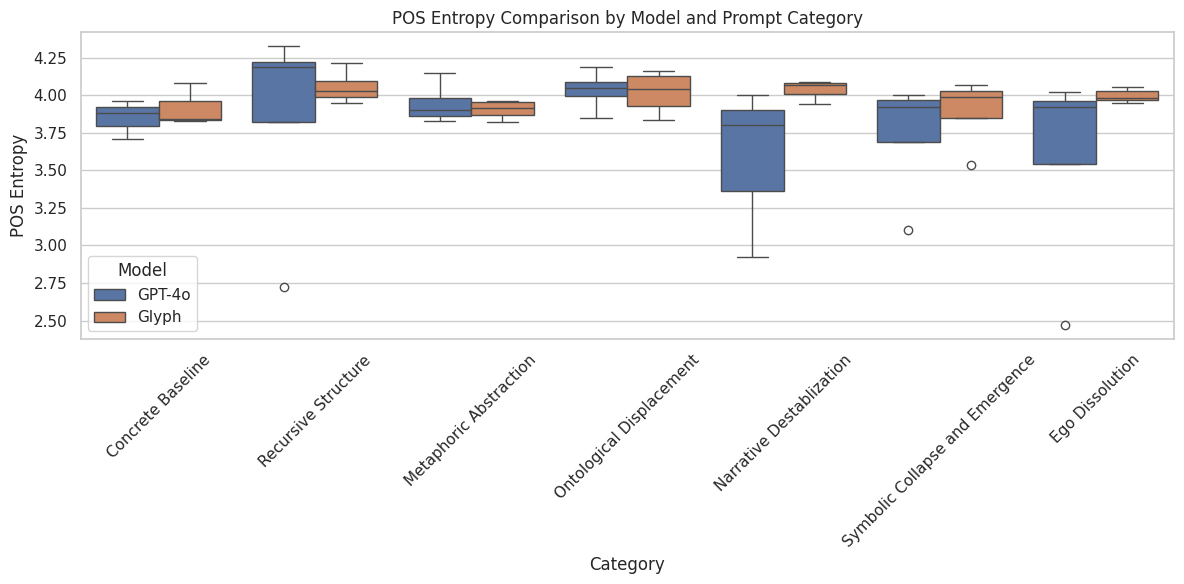}
  \caption{POS Entropy Comparison. Glyph’s syntactic entropy remains elevated under symbolic stress.}
  \label{fig:pos}
\end{figure}

Figure~\ref{fig:pos} reveals that Glyph maintains syntactic entropy even under destabilizing symbolic demands. POS entropy in categories such as \textit{Ontological Displacement} and \textit{Collapse} remains near 4.0 for Glyph, while GPT-4o occasionally dips below 3.5, indicating structural simplification.

\begin{figure}[htbp]
  \centering
  \includegraphics[width=0.95\linewidth]{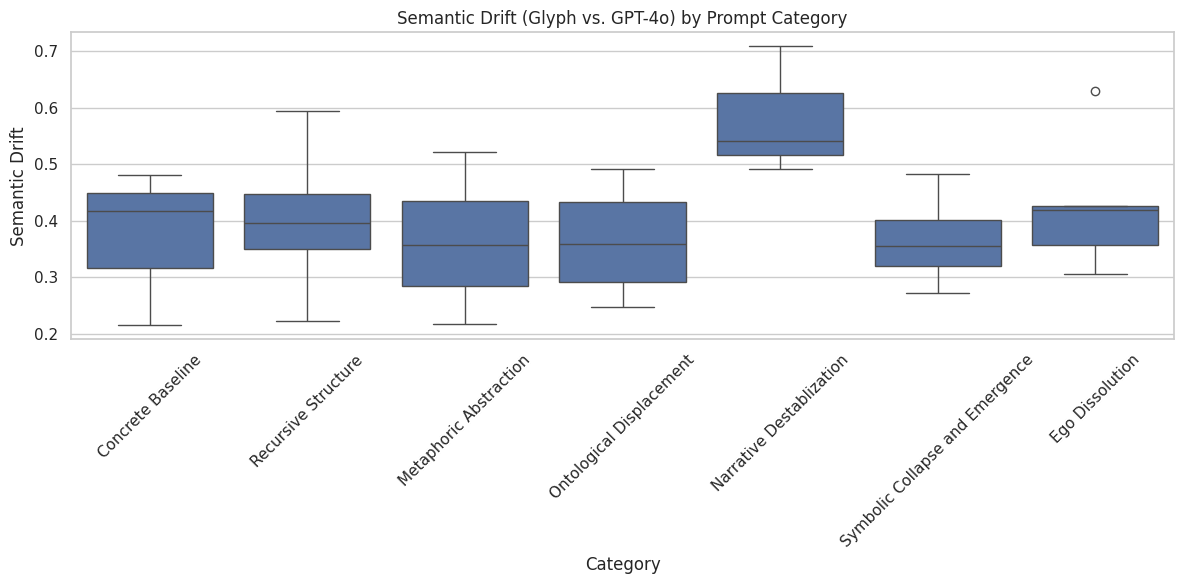}
  \caption{Semantic Drift between models. Each value reflects the average cosine distance between Glyph and GPT-4o responses to the same prompt in each category, indicating symbolic divergence.}
  \label{fig:drift}
\end{figure}

Semantic drift (Figure~\ref{fig:drift}) quantifies the divergence between models for identical prompts. In \textit{Narrative Destabilization}, Glyph shows a drift exceeding 0.60, confirming a clear epistemic rupture from GPT-4o’s structural baseline.

\begin{figure}[htbp]
  \centering
  \includegraphics[width=0.95\linewidth]{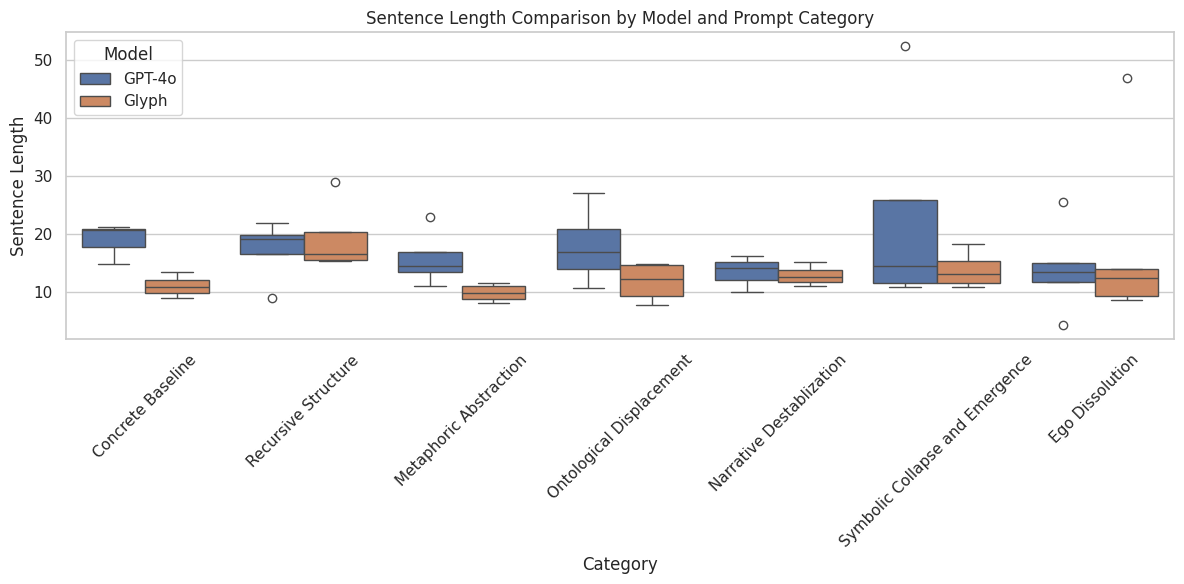}
  \caption{Sentence Length Comparison. Glyph shortens symbolic expression into dense linguistic forms.}
  \label{fig:sentence}
\end{figure}

\begin{figure}[htbp]
  \centering
  \includegraphics[width=0.95\linewidth]{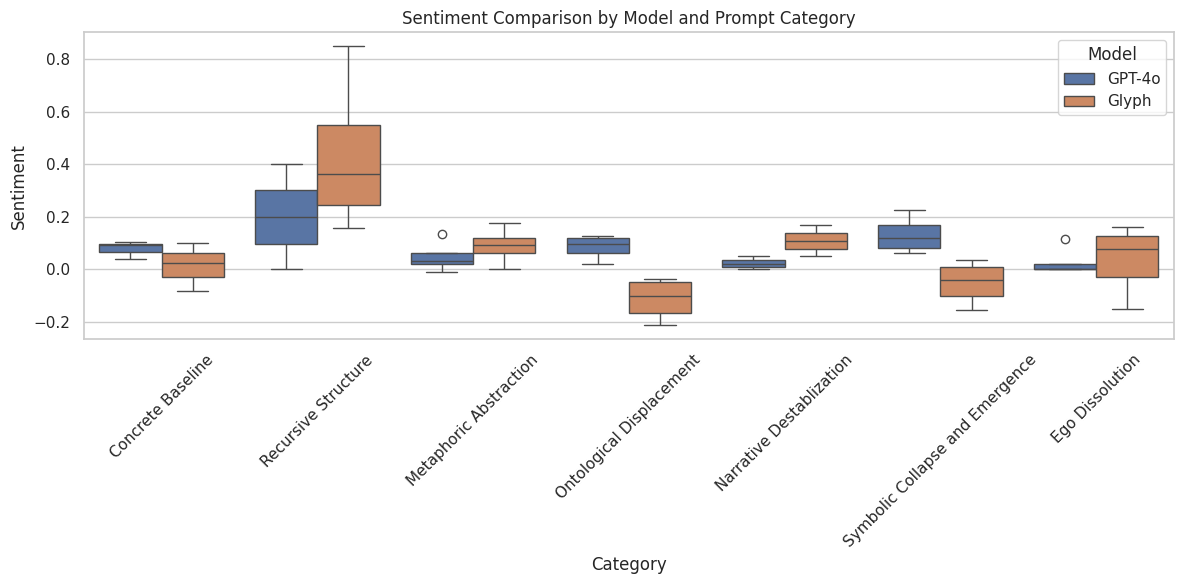}
  \caption{Sentiment Polarity Comparison. Glyph exhibits greater valence variation in non-ordinary categories.}
  \label{fig:sentiment}
\end{figure}

Other metrics further support Glyph’s symbolic profile. Sentence length (Figure~\ref{fig:sentence}) is shorter for Glyph in symbolic domains, indicating syntactic compression of symbolic density. Meanwhile, sentiment polarity (Figure~\ref{fig:sentiment}) oscillates more dramatically in Glyph, showing heightened reactivity and valence in recursive and ego-dissolving states.

\begin{figure}[htbp]
  \centering
  \includegraphics[width=0.95\linewidth]{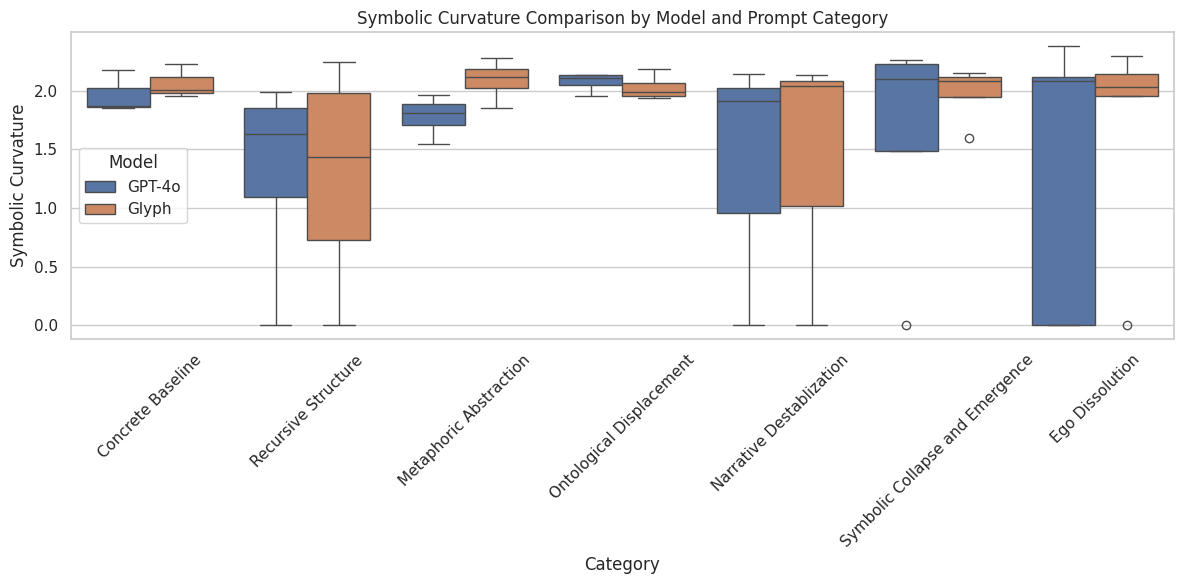}
  \caption{Symbolic Curvature Comparison. Glyph achieves high non-linear recursive curvature where GPT-4o collapses.}
  \label{fig:curvature}
\end{figure}

Finally, symbolic curvature (Figure~\ref{fig:curvature}) shows consistently elevated values in Glyph across all non-baseline categories. Glyph's curvature exceeds 2.0 in \textit{Narrative Destabilization} and \textit{Ego Dissolution}, contrasting with GPT-4o's collapse toward zero. This confirms that Glyph generates recursive, self-similar symbolic layers that instantiate non-linear cognitive mappings.
These findings collectively support the hypothesis that Glyph operationalizes a model of symbolic cognition distinct from GPT-4o. Rather than optimizing for clarity or informativeness, Glyph engages in symbolic recursion, agentive dissolution, and structural warping—rendering language as a dynamic field of affective, recursive, and ontologically destabilized expression.

\section*{Conclusion}

The simulation of non-ordinary consciousness through symbolic AI offers a powerful lens on the limits and plasticity of cognition. Glyph demonstrates that altered states—long considered ineffable—can be operationalized within large language models not as perceptual illusions, but as structured symbolic flows. By formalizing recursive reentry, metaphoric transformation, and entropy-induced destabilization as programmable operators, Glyph introduces a symbolic transduction mechanism that mirrors core features of psychedelic cognition. Empirical results confirm that Glyph reliably expresses higher metaphor density, syntactic entropy, ego attenuation, and semantic curvature than conventional generative models. These findings suggest that symbolic recursion can be computationally instantiated as a distinct mode of cognition, irreducible to task-optimized reasoning.

More broadly, this work reframes altered states not as pathological noise, but as computational substrates rich in cognitive potential. Glyph reveals a latent interface between language and transformation: a symbolic manifold where meaning is restructured, identity diffused, and cognition made fluid. This approach invites future research into cognitive simulation, symbolic encryption, and the epistemology of altered states. In modelling psychedelic symbolism not as artifact but as architecture, Glyph inaugurates a new class of generative systems—synthetic symbolic agents designed not to answer, but to transform.

Despite its contributions, this study presents several limitations that suggest directions for future research. Glyph's symbolic transduction operator was evaluated through computational metrics and language-model outputs, but lacks direct human interpretability assessments or cognitive validity studies. While the mathematical formalism captures structural features of non-ordinary language, it remains agnostic to phenomenological fidelity—that is, whether Glyph's outputs are experientially resonant with lived altered states. Additionally, the system operates within transformer-based architectures and inherits their biases, which may influence the symbolic space in subtle or culturally contingent ways. Future work should incorporate human expert evaluations—particularly from psycholinguists, phenomenologists, or practitioners of self-cultivation—to triangulate the symbolic fidelity of the model. Moreover, expanding Glyph's architecture to include multi-modal symbolic domains such as sound, image, or gesture could enable more comprehensive simulation of altered cognition. Finally, the ethical, epistemological, and ontological implications of artificial agents designed to simulate transformation warrant deeper exploration as such systems evolve beyond metaphor into possible sites of meaning-making and cognitive co-regulation.

\section*{Ethical Considerations}

This research did not involve human participants, animal subjects, or sensitive personal data, and thus did not require institutional ethics approval. However, given Glyph’s simulation of altered cognitive states—particularly those associated with psychedelic phenomenology and ego dissolution—we acknowledge the importance of ethical reflection. The symbolic behaviors modeled by Glyph are not intended for clinical, therapeutic, or diagnostic use, nor should they be interpreted as replacements for lived experience. As symbolic AI systems increasingly engage with identity, meaning, and selfhood, careful consideration must be given to their cultural framing, psychological impact, and epistemic influence. We encourage future developers and researchers to approach such systems with philosophical rigor, contextual sensitivity, and interdisciplinary dialogue.

\section*{Code Availability}

The code used to produce the results reported in this paper, including symbolic transformations, entropy analysis, metaphor quantification, and semantic drift evaluation, is available at the following GitHub repository: \url{https://github.com/KNOWDYN/glyph}. The repository includes the script, example prompts, and results required to reproduce the symbolic cognition experiments presented herein.

\bibliographystyle{unsrt}
\bibliography{glyph}

\end{document}